%% file: main.tex
\documentclass[aps,journal=prapplied, twocolumn]{revtex4-2}
\input{commands}
\begin{abstract}
Laser propulsion has been proposed for relativistic interstellar flights, but it faces the significant challenge of requiring extremely powerful laser radiation due to the inherently low momentum transfer between the beam and the sail.  The photon-recycling technique enhances thrust by transferring momentum through multiple reflections within a cavity setup, formed by the lightsail and a ground-based mirror in a laser system array. In this work, a delay differential model is developed to describe the evolution of the beam and thrust, incorporating both the Doppler effect and the round-trip time delay experienced by each beam component.
With optimized multilayer reflectors, the thrust performance gain is shown to be significant for interstellar flight, though limited by diffraction and the necessity of removing harmful redshifted radiation that could overheat the lightsail.
By balancing thrust performance with thermal stability, we derive a simple condition for determining the spectral requirements of the mirrors. Given a selected laser wavelength, this condition fully specifies the necessary properties of the cavity mirrors, enabling the same system to effectively support a range of launch protocols.
\end{abstract}

\begin{document}
\author{F. Lorenzi$^{*, 1,2}$, L. Salasnich$^{1,2,3,4}$ and M. G. Pelizzo$^{5, 6, 7}$}

\affiliation{
$^{1}$Dipartimento di Fisica e Astronomia “Galileo Galilei", Università di Padova, via Marzolo 8, 35131 Padova, Italy \\ 
$^{2}$Sezione di Padova, Istituto Nazionale di Fisica Nucleare (INFN), Via Marzolo 8, Padova, 35131, Italy \\
$^{3}$Istituto Nazionale di Ottica (INO), Università di Padova, via Nello Carrara 1, Sesto Fiorentino, 50019, Italy\\
$^{4}$ Padua QTech Center, Università di Padova, via Gradenigo 6A, Padova, 35131, Italy\\
$^{5}$Dipartimento di Ingegneria dell’Informazione, Università di Padova, via Gradenigo 6A, Padova, 35131, Italy\\
$^{6}$Istituto di Fotonica e Nanotecnologie, Consiglio Nazionale delle Ricerche, via Trasea, 7, Padova, 35131, Italy\\
$^{7}$Centro di Ateneo di Studi e Attività Spaziali “Giuseppe Colombo”, Università di Padova, via Venezia, 15, Padova, 35131, Italy\\
$^{*}$ Corresponding author, email: \texttt{francesco.lorenzi.2@phd.unipd.it}
}
\title{Optical Cavity in Relativistic Regime for Laser Propulsion}
\maketitle

\section{Introduction}
Interstellar exploration and relativistic flight are attracting growing scientific interest, particularly in the search for exoplanets and habitable environments beyond the Solar System. 
In the coming decades, a first lunar outpost will be established \cite{caluk2023,wingo2023,tomlinson2023}, 
which could serve as a center for scientific research and technological innovation, and possibly as a potential launch base for spacecrafts exploring other planets and deep space. 
At the moment, the most
effective thrusting techology for spacecrafts is provided by chemical propulsion, which, however,
requires transport and use of large quantities of polluting fuel, making it
less compatible with frequent trips.
A remarkable alternative is laser propulsion \cite{Phipps2024}, that may be implemented using thrust from the ablation of propellant situated at the spacecraft \cite{Rezunkov2021, Levchenko2018, Felicetti2013, Zhang2015, Phipps2010, Duplay2022}, or in a propellant-free way, where the thrust is provided by the momentum
exchange between the incident photons and the lightsail, is also receiving great
attention, as researchers aim to demonstrate that this clean technology can facilitate faster travel 
between planets within the solar system \cite{tung2022low, santi2023swarm} and achieve relativistic 
velocities to propel
nano-sats beyond the heliopause
\cite{Marx1966,Forward1984,lubin2016roadmap,kulkarni2018relativistic,lubin2020path, LubinP2022, 
Worden2021, Parkin2018}. 
\Ra{
Advancements in the technology of optical surfaces prospected the usage of metasurfaces \cite{salary2020photonic, siegel2019self, davoyan2021photonic, chang2024broadband} and graphene-enhancement \cite{Zhang2015} for lightsails.
}
Nevertheless, some currently limiting factors prevent the implementation of this
technique, hereafter indicated as Direct Energy Propulsion (DEP), in the short
term, as beam power needed to guarantee an efficient acceleration is on the order of tens of
GW, and the sail must be mechanically stable and very lightweight \cite{santi2022multilayers,lubin2020path, lubin2016roadmap, Worden2021, campbell2021relativistic, siegel2019self, gao2024dynamically, santi2022multilayers,santi2023swarm, myilswamy2020photonic, cassenti2020design, savu2022structural, Taghavi2022}.
One proposed solution involves the use of a kilometer-sized phased array capable of combining beams to achieve the necessary power levels. However, realizing such an array demands substantial technological advancements and significant resource investments \cite{lubin2016roadmap, lubin2020path}.

Although the transfer phase would last only a few minutes to a few hours, the associated energy consumption remains high. To address this, the ``photon recycling" technique has been proposed, with the aim of enhancing thrust efficiency without increasing the power source requirements \cite{kulkarni2018relativistic, bae2021photonic, bae2022photonic}. 
Although the transfer phase would last only a few minutes to a few hours, the associated energy consumption remains high. To address this, the ``photon recycling" technique has been proposed, with the aim of enhancing thrust efficiency without increasing the power source requirements \cite{kulkarni2018relativistic, bae2021photonic, bae2022photonic}. 

Theoretical studies on the dynamics of light in a cavity with a moving mirror are not new \cite{baranov1967electromagnetic,
moore1970quantum}; however, for the first time, this study presents a comprehensive theoretical model specifically tailored for 
the relativistic regime, enabling an accurate estimation of the final velocity and demonstrating the potential advantage of this technique compared to
DEP for relativistic travels.
A key limitation of earlier studies \cite{kulkarni2018relativistic, lubin2016roadmap, lubin2020path} resides in the adoption of a foundational corpuscolar model, where momentum transfer is calculated as the sum of contributions from all reflections without accounting for the delay of reflected photons; additionally, previous models neglect the Doppler frequency shift of the multiply-reflected radiation, treating all reflected components equivalently from a spectral perspective. In contrast, the present model addresses these limitations, resulting in a more accurate estimate of the thrust development in the relativistic regime.
This article is organized as follows: in Section II, we provide the theoretical model for the multiple reflection relativistic regime, in Section III we derive the condition for the reflection bandwidth, and in section IV we propose the multilayer designs for the reflectors, and test them in the case of a typical launch protocol.

\section{Theory of multiple reflection thrusting in the relativistic regime}
\begin{figure}
        \centering
        \includegraphics[width=0.93\linewidth]{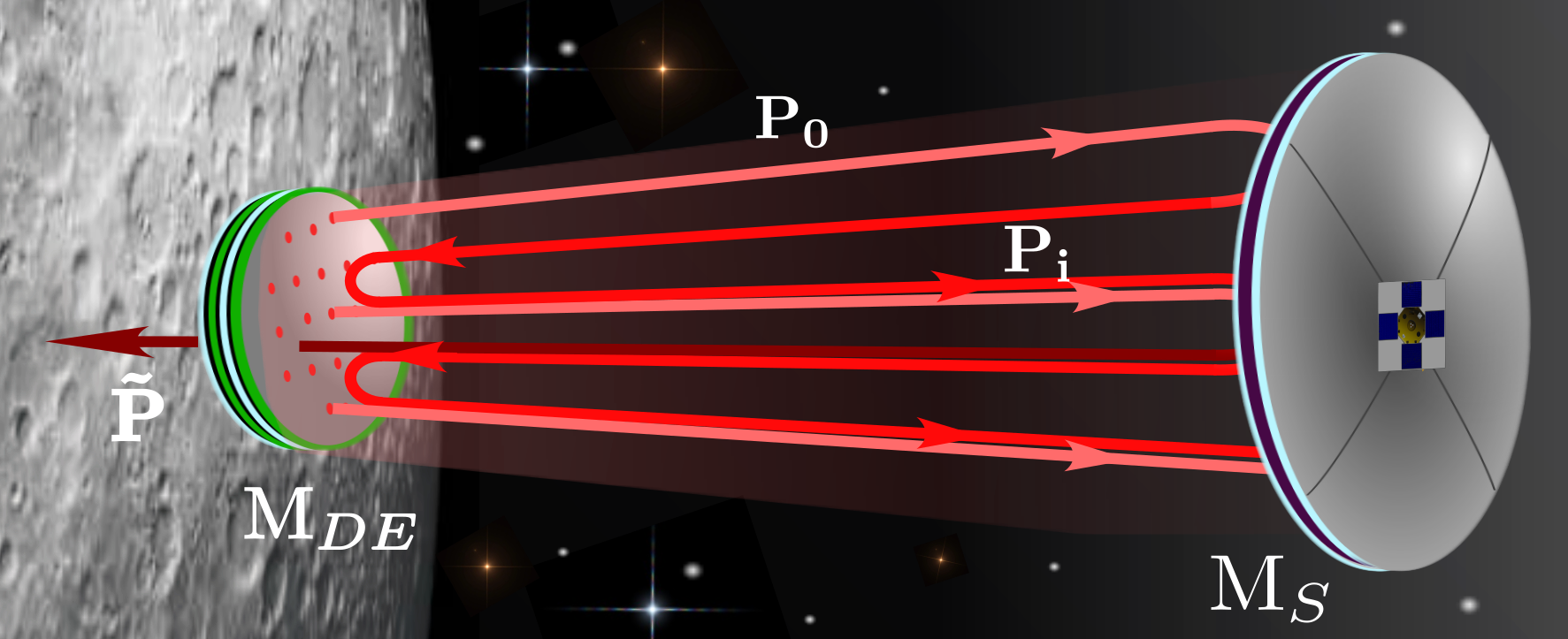}
        \caption{Layout of a Moon-based multiple reflection laser propulsion system. The colored lines represent the light originating from the laser array, denoted with $P_0$, and trapped in the cavity formed by the lightsail ${M}_{S}$ and the ground mirror, ${M}_{DE}$ getting progressively reshifted ($P_i$ is the $i$th reflection) until removal through transmissio, resulting in transmitted component $\tilde{P}$ at the cut-off threshold.}
        \label{fig:depiction}
\end{figure}

The concept of a photon recycling system is shown in Fig.~\ref{fig:depiction}a, where the reflective lightsail ($\mathrm{M}_S$) and a ground-based mirror ($\mathrm{M}_{DE}$), positioned at the laser site, form a cavity \cite{kulkarni2018relativistic, bae2021photonic, bae2022photonic}. In this setup, photons that bounce off the lightsail are redirected back toward it by the ground-based mirror, effectively recycling the photons to enhance thrust efficiency and reduce energy waste. To differentiate this system from traditional DEP, the propulsion system is hereafter referred to as Multiple-reflection Direct Energy Propulsion (MDEP).

The analysis of the dynamics leads to the definition of the requirements of the cavity mirrors in terms of the reflection bandwidth.
In the following, we introduce our theoretical assumptions and derive the delay differential equation model starting from the relativistic Newton equation in presence of radiation pressure.

\subsection{Relativistic equation of motion for the sail-radiation interaction}
Let the lightsail move along the $x$ axis, and let the spacecraft position in time be denoted by $q(t)$. The mirror $\mathrm{M}_{DE}$ at the laser source is placed at $x=0$, while the lightsail mirror $\mathrm{M}_S$'s position over time is denoted 
by $q(t)$, with a fixed initial position $q(0)=q_0>0$. In the approximation that the lightsail is a simple plane mirror, orthogonal to the beam propagation direction, the transverse force is negligible with respect to the $x$ component of the force, represented by the radiation pressure. 
The relativistic Newton equation in presence of only longitudinal force reads (see also Appendix A)
\begin{equation}\label{eq:newton}
    \gamma^3(t)\, m \, \ddot{q}(t) = F(t) \,,
\end{equation}
where $\gamma(t)=1/\sqrt{1-\beta(t)^2}$ is the relativistic Lorentz-Fitzgerald factor, $\beta(t) = \dot{q}(t)/c$ being the normalized velocity to the speed of light $c$, $m$ is the lightsail mass, and $F(t)$ the force acting on the sail. The force is exerted by the radiation pressure of a laser beam.
\Rb{
The power $P$ of the beam over the whole lightsail surface is computed integrating the intensity $I$ of the wave at each point of the lightsail. The intensity, in turn, is obtained taking the time average of the Poynting vector of the electromagnetic field in the following way.
We define the Poynting vector as $\mathbf{S} = \mathbf{E}\times\mathbf{H}$, with $\mathbf{E}$ and $\mathbf{H}$ the electric and magnetic field vectors. The intensity through a surface of normal vector $\hat{\mathbf{n}}$ is given as the time average, denoted by $\langle\cdot\rangle_t$, of the normal component of the Poynting vector  $I = \langle\mathbf{S}\cdot\hat{\mathbf{n}}\rangle_t$. Assuming a plane wave and normal incidence, it is possible to express the intensity as
\begin{equation}\label{intensity}
    I =  \varepsilon_0c\langle|\mathbf{E}|^2\rangle_t \,.
\end{equation}
The time average is considered over a long time compared to the period of the optical wave, but such that it can be still considered instantaneous with respect to the timescale of the motion of the mirror. Therefore, it is possible to write the expression of the force as a function of the time-varying power $P(t)$ of a narrowband beam over the lightsail surface as
\begin{equation}\label{eq:force}
    F(t) = \frac{(2r_1+\alpha)P(t)}{c} \ D(\dot{q}(t)) \,,
\end{equation}
where $r_1$ and $\alpha$ are the reflectivity of the sail and its absorptivity at the frequency of the radiation, assuming it monochromatic, }and $D(\dot{q}(t))$ is the Doppler shift coefficient.
The Doppler shift that affects the beam is the ratio between the impinging $\omega_i(t)$ and reflected $\omega_r(t)$ angular frequencies at the instant $t$ \Ra{(see Appendix B)}. Therefore the coefficient is given by
\begin{equation}\label{coffi}
D(\dot{q}(t)) = \frac{\omega_r(t)}{\omega_i(t)} = \frac{1-\beta(t)}{1+\beta(t)} \,.
\end{equation}
Combining \Rb{Eq.(\ref{eq:newton}, \ref{eq:force}, \ref{coffi})}, and isolating the acceleration $\ddot{q}(t)$, we get
\begin{equation}
{\ddot q}(t) = {(2r_1+\alpha)P(t) \over mc} \left( 1 - {{\dot q}(t)^2\over c^2} \right)^{3/2}
{1-{{\dot q}(t)\over c}\over 1+{{\dot q}(t)\over c}} \,,
\label{comeloro}
\end{equation}
in accordance \Ra{with } past literature \cite{kulkarni2018relativistic, lubin2016roadmap, ilic2018nanophotonic, atwater2018materials}. 

\subsection{Delay differential model for the multiple reflections}
\begin{figure}
        \centering
        \includegraphics[width=0.8\linewidth]{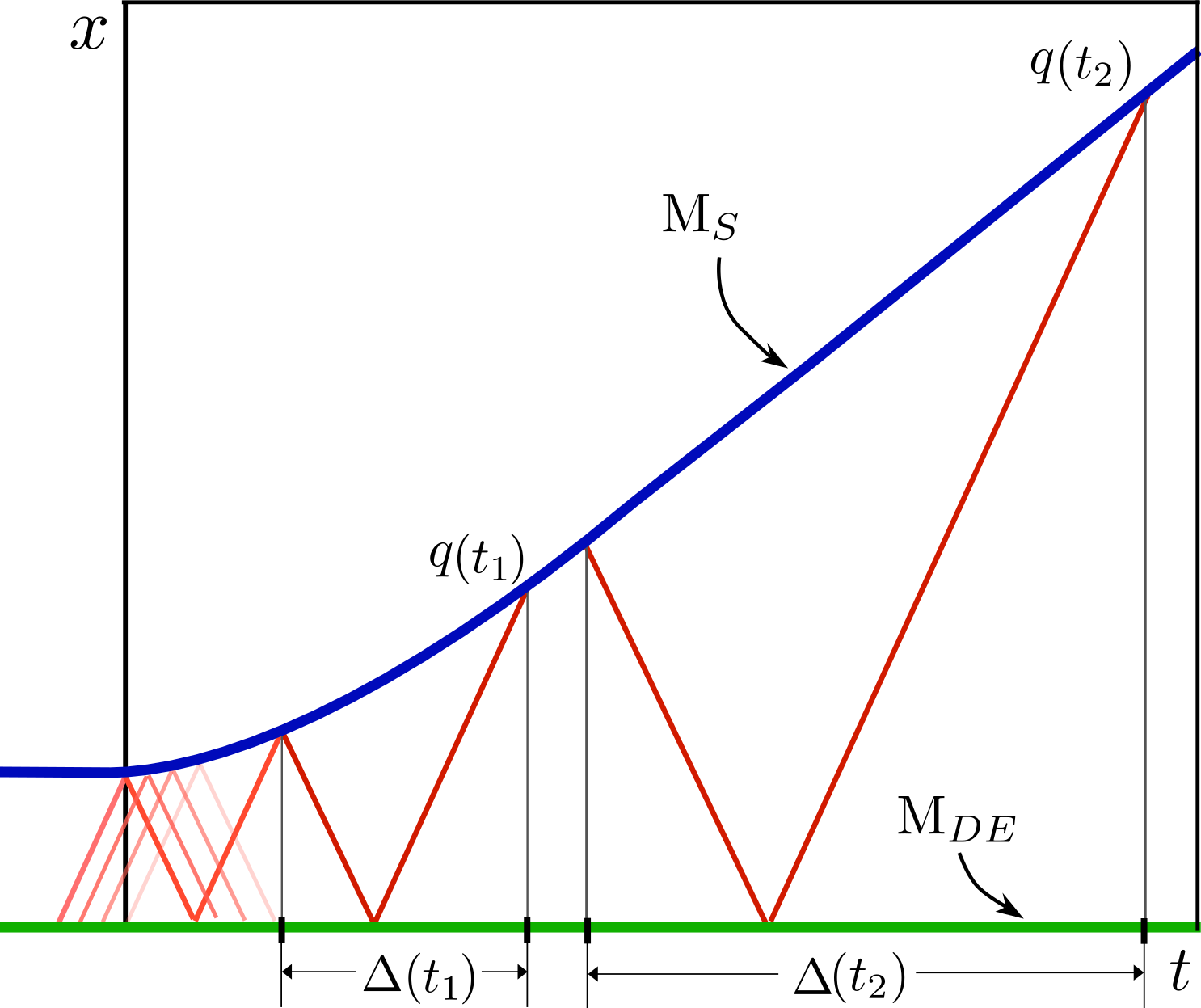}
        \caption{Illustration of the delay differential model used for the lightsail motion. The delay times at given times $t$ are computed with knowledge of the past trajectory of the sail, and they determine all the reflected radiation components acting on the sail recursively, by linking them to the ones present in the sail at the previous reflection time $t-\Delta(t)$.}
        \label{fig:theory}
\end{figure}
The equation of motion~(\ref{comeloro}) is usually utilized with a constant power $P(t)=P_0$ \cite{kulkarni2018relativistic}, \Ra{in which case it is possible to solve the equation by quadrature}, but in principle $P(t)$ can be the incident power of an arbitrary, time-dependent beam or set of beams, as long as the assumption that they propagate along the $x$ direction holds. In this work, a variable number of reflections is assumed to be present at the lightsail at a given time, due to the finite propagation time over long distances. The reflected radiations components are combining at the lightsail in an incoherent-like way.
\Ra{
Within the previously given definition of the intensity, Eq.~(\ref{intensity}), the intensity of a set of beams of different frequency, each with an electric field of $\mathbf{E}_j$, is considered by expressing the electric field as a sum of all the fields as follows 
\begin{equation}
    I = \varepsilon_0c\langle |\sum_j{\mathbf{E}_j}|^2\rangle_t = \varepsilon_0 c\sum_j \langle|\mathbf{E}_j|^2\rangle_t \,,
\end{equation}
using the fact that cross-terms vanish in the time average due to the fact that they have different frequencies. This shows that, within these simplifying assumptions, the intensity of such sum of beams is the sum of the intensities of each beam. For the purposes of the computation of the thrust, the beams may be considered incoherent-like, and they are weighted by the reflection and absorption coefficients at each respective frequency.
With this result, and introducing the frequency-dependent reflectivity and absorptivity, t}%
he total power ${P}(t)$ at the lightsail can be \Ra{subsituted with }  a time-dependent sum \Ra{of components}, with the number $N(t)$ increasing over time
\Rb{
\begin{equation}\label{powersum}
     (2r_1+\alpha){P}(t) \to \sum_{i=0}^{N(t)}(2r_1(\omega_i'(t))+\alpha(\omega_i'(t))){P}_i(t),
\end{equation}
}
where ${P}_i(t)$ is the power of the beam reflected $i$ times
from the $\mathrm{M}_S$, and that impinges on it at time $t$ \Rb{, and $\omega'_i(t)=\sqrt{D(\dot{q}(t))} \ \omega_i(t)$ its frequency in the frame of reference of the lightsail, calculated as the Doppler shift of $\omega_i(t)$, the frequency in the system of reference of the DE system. } 
The round-trip time of flight required for the wave to travel from the location of the 
precedent reflection to the actual location of the mirror at time $t$ is hereafter indicated with $\Delta(t)$, as depicted in Fig.~\ref{fig:depiction}fa. This latter 
quantity can be computed with knowledge of the full past trajectory of the lightsail, 
namely, it satisfies the relation:
\begin{equation}\label{eq:delta}
    c\Delta(t) = q(t-\Delta(t)) + q(t) \; .
\end{equation}

Let $\omega_0(t) = \omega_0$ be the frequency of the laser in the ground system, and 
${P}_0(t) = P_0$ its power.
At time $t$ the frequency $\omega_i(t)$ associated with the component $P_i(t)$ must be calculated using the Doppler function $D(\dot{q}(t))$.
In particular, the frequency $\omega_i(t)$ of the component $P_i(t)$ can be written for $i>0$ using a recursive relationship which involves $\omega_{i-1}(t)$ associated to the component $P_{i-1}(t)$:

\begin{equation}\label{ricorsioneomega}
    \omega_i(t) = D(\dot{q}(t-\Delta(t))) \, \omega_{i-1}(t-\Delta(t)) \qquad \text{if } t\geq t_i
\end{equation}

where $t_i$ is the time in which the $i$-th reflection arrives at the mirror $\mathrm{M}_S$. 
\Ra{
The physical meaning of Eq.~(\ref{ricorsioneomega}) is the following. Each component with index $i>0$ impinging on $\mathrm{M}_S$ at time $t$ is originated from the component with index $i-1$ upon a previous reflection on $\mathrm{M}_S$. After the reflection, which induces a Doppler shift, the component will travel mantaining the same frequency  to $\mathrm{M}_{DE}$, and then come back on $\mathrm{M}_{S}$, with a time of flight $\Delta$. Before $t_i$, the $i$-th component is not defined.
}
Clearly, the time of arrival of the $i$-th reflection $t_i$ is satisfying the formula $t_i - t_{i-1} = \Delta(t_i)$ for $i=1, 2,...$, recalling $t_0=0$. 
In our model, it is supposed that the laser is switched on at time $t=-q(0)/c$, so that the
first arrival of the light at the sail at time $t_0=0$ (see also Fig.~\ref{fig:theory}). 
Since only photons that lose energy to the
sail are considered, necessarily $\omega_i(t)<\omega_0$ for every $t>0$ and $i=1, 2,... \,
$. 
Moreover, since the number of photons is not increasing, a similar
relation holds for the powers ${P}_i(t)<P_0$ for every $t>0$ and $i=1, 2,
... \,$.

In the case the round-trip time of flight $\Delta(t)$ is small, Eq.~(\ref{eq:delta}) can be simplified by expanding in power series to the first order as
\begin{equation}
    \Delta(t) \approx \frac{1}{c}(2q(t)-\Delta(t)\dot{q}(t)) \, .
\end{equation}
The total number $N(t)$ of wave reflections at time $t$ is simply $N(t)=\int_0^t dt'
\sum_{i=0}^\infty \delta(t'-t_i)$.
By solving for $\Delta(t)$,  is obtained.
\begin{equation}\label{deltaeffettiva}
\Delta(t) \approx \frac{2q(t)}{c +\dot{q}(t)}.
\end{equation}
\Ra{The above approximation can be validated directly by testing it back against the numerically obtained trajectory. Within this study, it is found to hold. }
The power associated with each component depends on the reflectances $r_1(\omega)$ 
at $\mathrm{M}_S$, as computed in the lightsail reference system, and $r_2(\omega)$, at the mirror $\mathrm{M}_{DE}$ ground reference site; the round-trip reflectance can be thus represented by the function 
\begin{equation}\label{welativistico}
    r(\omega, \dot{q}) = r_1(\sqrt{D(\dot{q}(t))}  \omega) \ r_2(D(\dot{q}(t))\omega) \,.
\end{equation}
\Ra{
The above relation is subtle in the fact that the reflectance $r_1$ is computed using the frequency in the reference frame of the lightsail (see Appendix B). } 
The $P_i(t)$ component can be written using a recursive relationship \Ra{
similar to Eq.~(\ref{ricorsioneomega}). A remarkable difference with the frequency recursive Eq.~(\ref{ricorsioneomega}) is that the power, unlike the frequency, is not conserved over all the round trip to $\mathrm{M}_{DE}$ and back to $\mathrm{M}_S$. There is some power loss also on $\mathrm{M}_{DE}$ due to diffraction and non-unitary reflectivity. This is modeled in the following way
}
\begin{align}\label{ricorsionep}
    {P}_i(t) =& (1-\delta_{i, 0})\Gamma_i(t-\Delta(t)) {P}_{i-1}(t-\Delta(t))\Theta(t-t_i) \nonumber  \\ &+ \delta_{i, 0}P_0 \,,
\end{align}
where $\Theta(t)$ is the Heaviside step function, $\delta_{i, 0}$ is the Kronecker delta, and 
\begin{equation}\label{eq:gamma-minestra}
  \Gamma_i(t) = r(\omega_{i-1}(t), \dot{q}(t)) \cdot D(\dot{q}(t)) \cdot f_i(q(t)) \,,
\end{equation}
is a round-trip reflection efficiency factor
\Ra{
including the diffraction coefficient $f_i(q(t))$, which represents the power loss of the $i$-th component due to the fact that the beam falls partially outside the mirror edges due to diffraction. For each component $i$, constituting a beam, the coefficient is computed as the ratio between the beam power impinging on the mirror at the $i+1$-th reflection and the initial power of the component. For $i=0$, the component is generated at $\mathrm{M}_{DE}$, whereas for $i\neq 0$, it is generated after the $i$-th reflection with the mirror (see also Fig.~\ref{fig:theory}). It follows that the $i=0$ component is diffracted only once since it only propagates forward and reflect on $\mathrm{M}_{S}$. On the other hand, the components with $i \neq 0$ undergo two diffraction processes, first, propagating backward from $\mathrm{M}_{S}$ to $\mathrm{M}_{DE}$,  and then, propagating forward from $\mathrm{M}_{DE}$ to $\mathrm{M}_{S}$. After reflecting on $\mathrm{M}_{S}$, the component gets relabeled with the label $i+1$. In the following, the fraction of power lost due to diffraction, dependent on the distance traveled by the light and its frequency, is denoted by $F_{\uparrow}(q, \omega)$ for the forward case, and as $F_\downarrow(q, \omega)$ for the backward case.  By considering the correct reflection instants, the coefficient reads
\begin{align}
    &f_i(q(t)) = \delta_{i, 0} \ F_{\uparrow}(q(t), \omega_0) \nonumber\\
    &+ (1-\delta_{i, 0}) F_{\uparrow}(q(t), \omega_{i-1}(t)) F_{\downarrow}(q(t-\Delta), \omega_{i-1}(t)) \,,
\end{align}
}

\Ra{
The forward and backward diffraction losses can be computed within the assumption that the beams can be modeled as Gaussian beams. If $w_\uparrow(q, \omega)$ is the beam width when propagated from $\mathrm{M}_{DE}$ to $\mathrm{M}_S$, and $w_\downarrow(q, \omega)$ in the opposite direction, the corresponding power losses are
\begin{equation}
    F_{\uparrow}(q, \omega) = 1-\exp\left[-\frac{1}{2}\left(\frac{d_S}{w_\uparrow(q, \omega)}\right)^2\right] \,,
\end{equation}
and
\begin{equation}
    F_{\downarrow}(q, \omega) = 1-\exp\left[-\frac{1}{2}\left(\frac{d_{DE}}{w_\downarrow(q,\omega)}\right)^2\right] \,.
\end{equation}

The beam emitted by the laser system, as well as the one reflected by the mirrors, can be approximated as a Gaussian beam if the size of the systems, considered to be circular, is sufficiently large with respect to the beam width. The ratio between the system radius and the beam radius is $\varphi$. With a value of $\varphi$ as low as  $\varphi=2$, the approximation is good, as the beam power out of the system is just \SI{3.35e-4}{} times the total power, and the output beam may be well considered a Gaussian beam.
This approximation is only valid for beams arriving at the mirrors with small diffraction, and needs in principle to be corrected with beam truncation effects. However, it is verified that in our case, all the beams requiring this kind of correction have very small power and therefore they can be neglected.

The Gaussian beam can be set to focus at a given distance by setting the relative phase of the phased array elements of the laser, or by curving the mirror surface. For the laser, this is possible only with very strict coherence requirements, which are still difficult to meet in practice, but recent work point out significant technological progress \cite{lubin2020path}. 
The spatial inhomogeneities over a very large mirror and array will degrade the beam quality, but there is the possibility of partially correct them in the first phases of the launch using adaptive optics and the aberration information carried by the backscattered beam.
\begin{figure}
        \centering
        \includegraphics[width=0.95\linewidth]{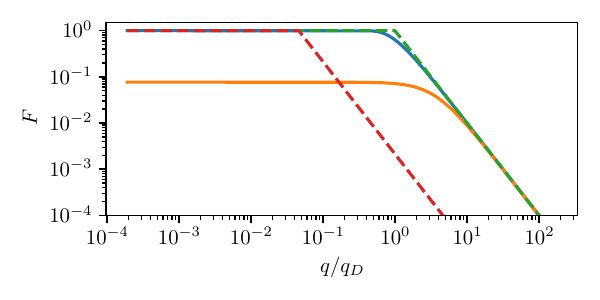}
        \caption{
        \Ra{Diffraction factors $F_\uparrow$ and $F_\downarrow$ as functions of the distance $q/q_D$ normalized to the diffraction distance $q_D= \sqrt{2}\pi \varphi \ d_{DE}d_{S}/\lambda_0$, with the width ratio $\varphi=2$.  The solid blue line is $F_\uparrow=F_{\downarrow}$ with optimal focusing. The dashed red line is the two-regimes diffraction function from Ref.~\cite{kulkarni2018relativistic}, i.e. in the case of uniform illumination of the aperture, and the dashed green line corresponds to the same two-regimes model rescaled for the Gaussian beams. The orange line is $F_\uparrow$ in absence of refocusing, when $d_{DE}= 100 \ d_{S}$.
        }
        }
        \label{fig:diffraction}
\end{figure}
The MDEP may take advantage from an active focusing of the mirrors and laser system, designed to keep the diffraction losses to a minimum while the lightsail travels along the optical axis. A Gaussian beam with the waist at $q_0$ of radius $w_0$ satisfies the radius relation $w(q) = w_0\sqrt{1+((q-q_0)/q_R)^2}$ with $q_R = \frac{\pi w_0^2}{\lambda}$ the Rayleigh range.
It is possible to show that there is a minimum spot size that can be obtained at a distance $q$ from an initial position where the beam radius is specified. Supposing that the sail and the mirror are capable of optimal focusing, the spot sizes can be written as 
\begin{equation}\label{eq:fwd}
    w_{\uparrow}(q, \omega) = \frac{\lambda q}{\pi \frac{d_{DE}}{2\varphi}} = \frac{4\varphi c  \ q}{\omega \ d_{DE}}\,,
\end{equation}
and 
\begin{equation}\label{eq:diffraction}
    w_{\downarrow}(q, \omega) = \frac{\lambda q}{\pi \frac{d_{S}}{2\varphi}} =  \frac{4\varphi c  \ q}{\omega \ d_{S}} \,,
\end{equation}
with this result, it is verified that $F_\uparrow(q, \omega) = F_\downarrow(q, \omega)$. 
When the spot sizes are much smaller than the dimension of $\mathrm{M}_S$ or $\mathrm{M}_{DE}$, a more defocused beam is preferrable to achieve a more homogeneous power distribution across the mirrors, and this can be achieved without major changes to the analysis of the power lost because of diffraction. 
In Fig.~\ref{fig:diffraction}, the diffraction coefficients (blue solid line) are represented and compared to the one of the previous diffraction model \cite{kulkarni2018relativistic} (red dashed line), which utilizes a function of the form $f(q, \omega)=\min\left\{(d_{DE}d_{S} \ \omega/(4\pi \cdot 1.22 \ c \ q))^2, 1\right\}$. They share the same qualitative behaviour, subdivided in two regimes, but with a rescaled transition distance. The case of lack of optimal focusing is also represented (orange line) with the forward diffraction factor in the case of $d_{DE}=100 \ d_{S}$, showing major losses for any value of the distance, due to the lack of convergence of the beam at the lightsail. The resulting diffraction factor is much smaller than one obtained with optimal focused widths even for asymptotically small distances (the value is less than 10\%), and suggests the importance of the correct focusing.

The diffraction model presented is an optimistic model, and can be used to calculate the absolute best performance that is obtainable in presence of the diffraction with Gaussian beams.
This model requires dynamically changing curvatures, whereas in a typical launch scenario, the curvature of the sail is difficult to reconfigure over time.
}

It should be noted that the delay differential model presented above reduces to the foundational corpuscular model described in \cite{kulkarni2018relativistic} under appropriate limiting conditions. In fact, if the Doppler shift is neglected, the frequency of each component in Eq.~(\ref{ricorsioneomega}) will stay the same. In this case, the power of each component can be expressed by the simplified equation ${P_i}(t) = r{P_{i-1}}(t-\Delta(t))$, where $r$ is the round-trip reflectance at the laser
frequency. Finally, if $\Delta \to 0$, the total power, as given in Eq.~(\ref{powersum}), is calcultated as ${P}(t) \to
P_\mathrm{inst}=  P_0/(1-r)$, in agreement with \cite{kulkarni2018relativistic}.
Within the approximations $r=1$ and $f=1$, solutions of Eq.~(\ref{ricorsionep}) depend solely on the initial condition $q_0$, as $\dot{q}(0)=0$, and they are obtained for two different values of $q_0$.
The joint solution of Eqs. (\ref{comeloro}-\ref{deltaeffettiva}) is a problem in the form of a state-dependent delay differential equation \cite{bellen2013numerical}, that is, for every time instant, the acceleration of the sail contains a dependence on the power present at a previous time, where this time instant is itself dependent on the previous power evolution.

\begin{figure}
    \centering
    \includegraphics[width=0.99\columnwidth]{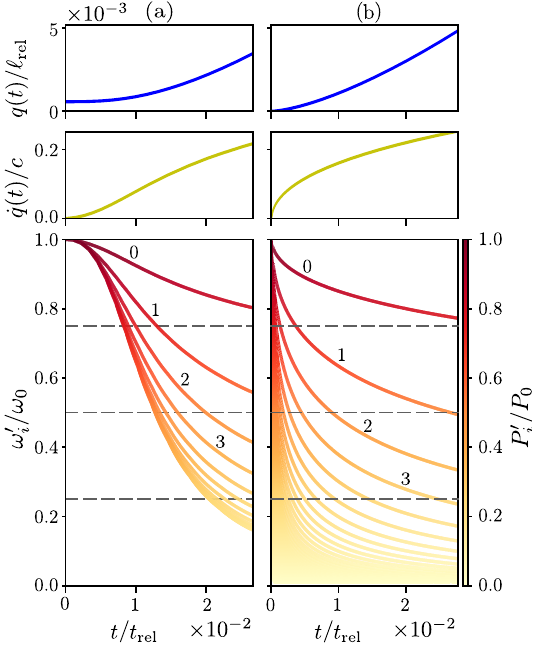}
    \vspace{0.2cm}
    \caption{Dynamics of the lightsail and the light spectrum over time, in the case of ideal reflectors. Top panels are the sail position $q(t)/\ell_\mathrm{rel}$; mid panels are the lightsail speed $\dot{q}(t)/c$;  bottom panels are the light spectra. In (a) the initial position is $q_0 = 5\times 10^{-4}\ell_{\mathrm{rel}}$, approximatively corresponding to a launch of a $m=$\SI{1}{g} sail from a distance of \SI{4e5}{km} with a power of $P_0=$\SI{50}{GW}. In (b) the initial position is $q_0 = 5 \times 10^{-7} \ell_{\mathrm{rel}}$, corresponding to a launch of a $m=$\SI{10}{g} sail from a distance of \SI{8e2}{km} with a power of $P_0=$\SI{20}{GW}. 
    The \Ra{dashed lines correspond to the frequencies above which, for all the time evolution, the power of each components is at least a given fraction of the initial power. From top to bottom, the fractions are $3/4$, $1/2$ and $1/4$ of $P_0$, respectively, and coincide with the frequency fraction in the vertical axis. The first three reflections are numbered. The solutions are expressed in terms of the relativistic characteristic time, $t_{\mathrm{rel}}=mc^2/P_0$, and the relativistic characteristic length, $\ell_{\mathrm{rel}}=mc^3/P_0$. }}
    \label{spectra}
\end{figure}

The solution of the differential equation is carried out numerically propagating in time with fourth-order Runge-Kutta technique, supplemented by a dynamic programming table, where the state-dependent delay serves as an index to set the interpolation of past values of ${P}_i(t)$.

The solutions are expressed in terms of the relativistic characteristic time, $t_{\mathrm{rel}}=mc^2/P_0$, and the relativistic characteristic length, $\ell_{\mathrm{rel}}=mc^3/P_0$. 
The power spectra over time are represented in Fig.~\ref{spectra}, panels (a) and (b) for the two selected cases with different initial $q_0 / \ell_{\mathrm{rel}}$ values. The notation $\omega_i^{\prime}$ and $P_i^{\prime}$ indicates that the frequencies and powers are calculated in the lightsail reference frame, which is instrumental in defining the requirements for the lightsail. A large number of components are generated at the initial stage of the dynamics, while when the lightsail speeds up, the time interval between new components increases. The shadowed region corresponds to the band where the amount of power resulting from each reflected component is at least $50\%$ of the initial power $P_0$. The relative frequency corresponding to this criterion is independent from the launch configuration, as it can be seen by comparing the two bottom plots of Fig.~\ref{spectra}. This property, valid for any power ratio, is independent of the initial position $q_0$ and reflects the fact that, with ideal reflectors, the power of each component scales proportionally to its frequency due to photon conservation. The same property holds when computing the radiation at $\mathrm{M}_\mathrm{DE}$. Therefore, within given approximations, the reflector requirements will be independent of the launch initial conditions.

On the other hand, in the present model the dynamics is strongly dependent on the initial condition, which determines the effectiveness of the multiple reflections approach. When the sail's initial position is far away, the power gain is severely hindered by Doppler shift and time delay, as shown in Fig.~\ref{spectra}a, while this does not happen for a much smaller initial condition, represented in Fig.~\ref{spectra}b. The upper portions of Fig.~\ref{spectra}a and Fig.~\ref{spectra}b show the lightsail normalized position $q(t)/\ell_\mathrm{rel}$ over time, from which it can be seen that the second launch configuration (b) spans a higher length before the end of the launch. This is due to the faster initial acceleration, as shown in the velocity $\dot{q}(t)/c$ plots in the middle panels. The faster acceleration of the lightsail in the case (b) is linked to the more efficient reflection process, shown in the lower panels, as many reflections accumulate in a short time, compared to case (a). This effect is also remarkably appreciated in comparing our predictions to the foundational model \cite{kulkarni2018relativistic}. In such a model, the calculations in a case with ideal reflectors would result in a divergent summation for the power, so a finite but frequency independent round trip reflection is assumed.

\begin{figure}
    \centering
    \includegraphics[width=\columnwidth]{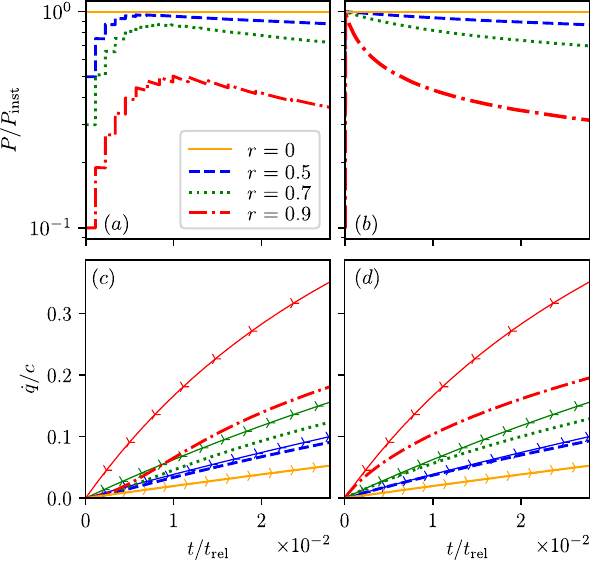}
    \caption{\Ra{Evolution of power and velocity over time. Panels (a) and (b) represent the } ratio between the power at the sail surface $P$ as predicted with the present model, and $P_{\mathrm{inst}}$ as given by the foundational \Ra{model of Ref. \cite{kulkarni2018relativistic}}, for the same \Ra{round-trip } reflectance.  \Ra{Panels (c) and (d) represent the final velocity with the same range of reflectances, including also the numerical solution of the instantaneous model (solid lines) and its analytical solution as a benchmark for the numerical method (stars). Initial conditions of the panels (a) and (c)  are the same of panel (a) of Fig.~\ref{spectra}, and the ones of panels (b) and (d) are the same of panel (b) of Fig.~\ref{spectra}. }\Ra{Since in panels (a) and (b) $P/P_\mathrm{inst}<1$ for the whole dynamics, the traditional model systematically overestimates received power. 
    }}
    \label{main-point}
\end{figure}%

In Fig.~\ref{main-point} the ratio of the total power predicted within our model and $P_{\mathrm{inst}}$ of the foundational model is plotted for different frequency-independent round-trip reflectance values, and keeping $r_1=1$. It is interesting to notice that the total power always stays below the levels predicted by the previous model, and its evolution over time is nontrivial. The decay of the power for long times, boh in panel (a), and panel (b) is due to the Doppler shift of reflected radiation. The discontinuities clearly visible in panel (a) mark the arrival of various reflection components over time, while in panel (b) these are not visible as they are compressed at the very beginning of the launch. For low starting distances, a higher efficiency of the multiple reflection process is obtained, and the power tends to the maximum value before decaying.
\Ra{
In Fig.~\ref{main-point} panels (c) and (d), a comparison is performed between the previous foundational model for the multiple reflections, with its analytical solution, and the present theory, by plotting the lightsail velocity over time.
}
\section{Requirements on the reflection bandwidth} 
\begin{figure}
    \centering
    \includegraphics[width=\linewidth]{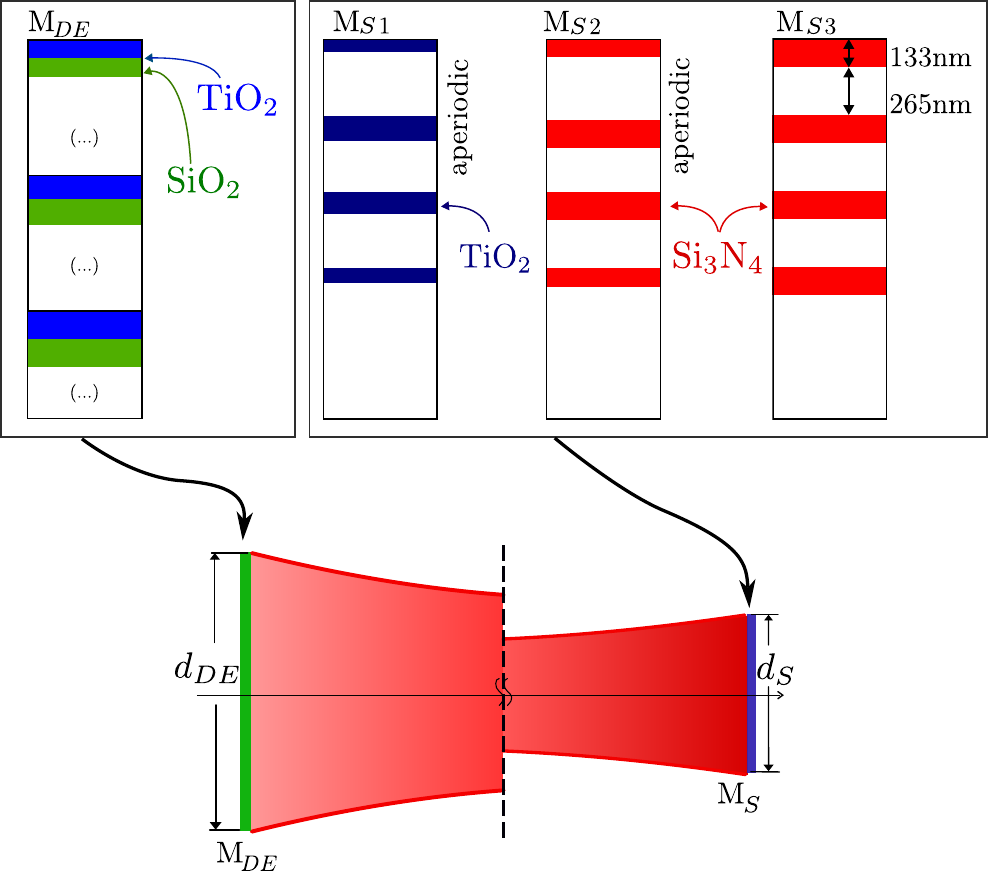}
    \caption{Schematic representation of the mirrors chosen for the thruster. The ground-based mirror $\mathrm{M}_{DE}$ is made of an aperiodic stack of silicon dioxide and titanium dioxide, whereas the lightsail mirrors $\mathrm{M}_{S1}$, $\mathrm{M}_{S2}$ and $\mathrm{M}_{S3}$ are made of and an aerogel-like material with unity index of refraction and density alike the one of air in standard temperature and pressure.}
    \label{fig:whereami}
\end{figure}
\begin{figure}
    \centering
    \includegraphics[width=0.9\columnwidth]{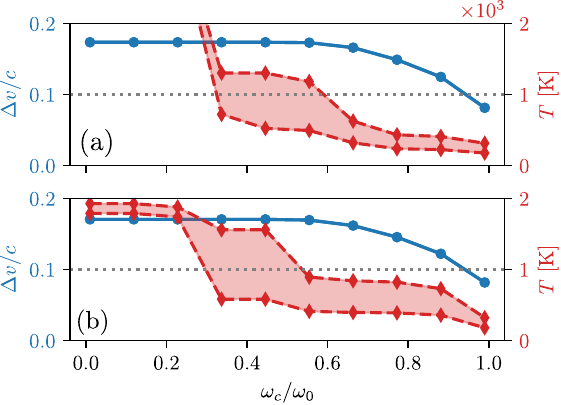}
    \caption{\Ra{Final velocity $\Delta v /c$ and maximum temperature $T$ as a function of the $\mathrm{M}_{DE}$ cutoff frequency $\omega_c$. Panel (a) and (b) use the same parameters as Fig.~\ref{spectra} panels (a) and (b), respectively. The sail diameter is set to $d_S=10$\SI{}{m}, and the thrust power is $P_0 = $\SI{50}{GW}. }}
    \label{fig:thermal}
\end{figure}%
\Ra{
During the thrust phase, the temperature of the lightsail must remain at least below the thermal stability limit of its constituent materials. In this section, a thermal model is developed and solved in conjunction with the previously discussed mechanical model. This is shown to provide design criteria for the whole system, where the interplay of the absorption and reflection and the band choices are critical. Since on $\mathrm{M}_{DE}$ the space and weight practical constraints are very flexible, and actively cooled reflectors can be mounted, this subsystem not considered thermally critical; consequently, the thermal analysis focus solely to the sail.

While previous models account for the full thermal dynamics of the lightsail as a lumped body \cite{santi2022multilayers}, an instantaneous model can be used to estimate its maximum temperature. The instantaneous model assumes the instant-by-instant equilibrium of heat absorbed into the sail and heat dissipated through thermal radiation. This assumption  is justified by considering the typical magnitudes of parameters involved in a interstellar launch case (see also Table~\ref{parameters} below): the typically slow change in absorbed power    and the very low thermal capacity of the sail, lead to an equilibration time constant which is orders of magnitude smaller than the timescale of the dynamics of the lightsail \cite{lin2025photonic}. 
In the present case, this is not directly valid for all time instants, since the absorbed power is discontinuous in time, due to the arrival of the components (an aspect clearly visible in Fig.~\ref{main-point}a). However, the discontinuity is always positive, and after the arrival of each component the incoming power stabilizes to a value that changes gradually. The equilibrium thermal model is expected to provide a good estimate of the maximum temperature under these conditions.
For any time $t$ the following equilibrium condition is
\begin{equation}\label{SB}
\sum_i P_i(t) a(\omega_i) = \frac{\pi^2}{2} d_{S}^2\int_0^\infty d\omega \  \varepsilon(\omega) I(\omega) \,,
\end{equation}
where $a(\omega)$ and $\varepsilon(\omega)$ are respectively the absorptance and the emittance at frequency $\omega$, and $I(\omega)$ is the Planck spectrum
\begin{equation}
I(\omega) = \frac{\hbar}{4\pi^3 c^2} \frac{\omega^3}{\exp\left(\hbar \omega/(k_BT)\right)- 1 } \,,
\end{equation}
where $\hbar$ is the Planck constant and $k_B$ is the Boltzmann constant. As in Ref.~\cite{tung2022low}, spherical emission is assumed.

To achieve sufficient heat dissipation, it is necessary to design a lightsail that is absorptive enough outside the reflection band \cite{atwater2018materials}.
Lightsail components with very high IR absorptance can be realized by using specialized nanostructures like the one proposed in Ref.~\cite{tung2022low}. The absorbance can be as high as  $0.5$ or greater out of the reflection band of the lightsail reflector. In the following, the previously proposed emissive nanostructure made of Si$_3$N$_4$ \cite{tung2022low} is considered as a benchmark for emission efficiency. The system is evaluated with the emissive structure attached to the back side of the lightsail reflector.
To model the absorptance of the emissive structure in the band of reflectance, we extrapolate it by approximating the structure as a slab with an averaged thickness. 
Since it distant from the operating wavelengths for which the structure was designed, this approximation is useful to compute the correct order of magnitude of the thermal effects. 
Furthermore, a factor of 10 lower absorptance is considered as a reference for a modified absorptance to be obtained through material engineering.
The overall absorptance is
\begin{equation}
    a(\omega) = a_S(\omega) +  (1-r(\omega)) \ a_E(\omega) \,,
\end{equation}
where $a_S$ is the absorptance of $\mathrm{M}_{S}$, and $a_E$ the one of the emissive structure. Considering the reflector materials to be low-loss in the reflection band, but assuming an imperfect reflectance due to the design constraints of $\mathrm{M}_S$, the approximation $(1-r(\omega))a_E(\omega) \gg a_S(\omega)$ is taken.

A well-known fact for multilayer reflectors is that, when keeping the complexity of the structure constant, high reflectance can be achieved only when reducing the bandwidth. By considing the design of $\mathrm{M}_{S}$ with multilayers as reported pictorially in Fig.~\ref{fig:whereami}, a reasonable bandwidth for the present application is of $1$\SI{}{\micro m} considering a fundamental frequency of about $1$\SI{}{\micro m}, typical of high power laser systems.
Therefore, it is appropriate to suppose that $\mathrm{M}_S$ has a reflection band from $\omega_0$, the laser frequency, to $\omega_0/2$. The $\mathrm{M}_{DE}$ cutoff frequrency $\omega_c$ can be optimized within this assumptions in the following way. Considering a simplified but representative model for the reflectance and absorptivity of $\mathrm{M}_S$, the mechanical and thermal models are solved together and the maximum temperature is found by considering a thrusting scenario under the same conditions of the simulation shown in Fig.~\ref{spectra}, and with a sail diameter of $d_S=10$\SI{}{m}, at a varying cutoff frequency $\omega_c$, in such a way that $\mathrm{M}_{DE}$ is modeled as a high-pass filter with this cutoff frequency. 
The results are shown in Fig.~\ref{fig:thermal}, and show that the final velocity is not increased by lowering the cutoff below about $\omega_0/2$. At the same time, the temperature is rising above \SI{1000}{K} for values lower than that, due to the backreflection of components in the mid-IR towards the sail. The value of \SI{1000}{K} can be taken as a reference value for the stability of bulk materials involved in the lightsail \cite{ccetin2022evaluation,liu2025thermal,zhang2023resilient,jiang2002compressibility,rocha2019environmental}. Based on all of the above considerations, the design value of the cutoff frequency is set to the reference value $\omega_c = \omega_0/2$. This corresponds approximately to the elimination of all the components with $\omega_0/2$ in Fig.~\ref{spectra}. Using the results obtained in the previous section in the case of perfect reflectivity, each one of them is expected to carry at most a power of $P_0/2$.

}

\begin{table*}
    \centering
    \begin{tabular}{l|c|l|l}
    \hline
    \hline
    \textbf{Design} & \textbf{Materials} & \textbf{Layer Thickness} & \textbf{Surface density} \\
    \hline
    $\mathrm{M}_{S1}$ & TiO$_2$ / Air-like& $[77, 335, 118, 329, 120, 349,
69, 330]$nm & \SI{1.633}{g/m^2}\\
    $\mathrm{M}_{S2}$ & Si$_3$N$_4$ / Air- like & $[126, 343, 158, 330, 160, 333,
132, 392]$nm & \SI{1.384}{g/m^2} \\
    $\mathrm{M}_{S3}$ & Si$_3$N$_4$ / Air- like & $4\times[133, 265]$nm & \SI{1.278}{g/m^2} \\
    \hline
    \hline
    \end{tabular}
    \caption{Lightsail mirror designs.}
    \label{tab:designs}
\end{table*}

\section{Thrust evaluation in the interstellar flight case}

The results obtained serve as a criterion for the actual design of the $\mathrm{M}_S$ and $\mathrm{M}_{DE}$ mirror structures. As previously discussed, \Ra{the cutoff frequency is set to $\omega_0/2$. } Lower frequencies, in fact, are not only less efficient, but also pose challenges for the thermal management of the lightsail, as they may fall within the absorption band of the lightsail materials. In general, the design of the lightsail include materials with high emissivity in mid-IR \cite{santi2022multilayers, jin2022laser, ilic2018nanophotonic, Feng2022, Brewer2022, Gao2022, Holdman2021}. It was demonstrated that the threshold is independent of the launch parameters and depends solely on the fundamental laser frequency:
in the present case the cavity is designed for a Nd:YAG laser operating at a wavelength of \SI{1064}{nm}. In this case, for the MDEP to be effective, $\mathrm{M}_S$ and $\mathrm{M}_{DE}$ need to exhibit high reflectivity in the wavelength range from $\lambda_0=\SI{1064}{nm}$ to $2\lambda_0=\SI{2128}{nm}$, while $\mathrm{M}_{DE}$ must filter out wavelengths greater than $2\lambda_0$. 

\begin{figure}
    \centering
    \includegraphics[width=0.75\linewidth]{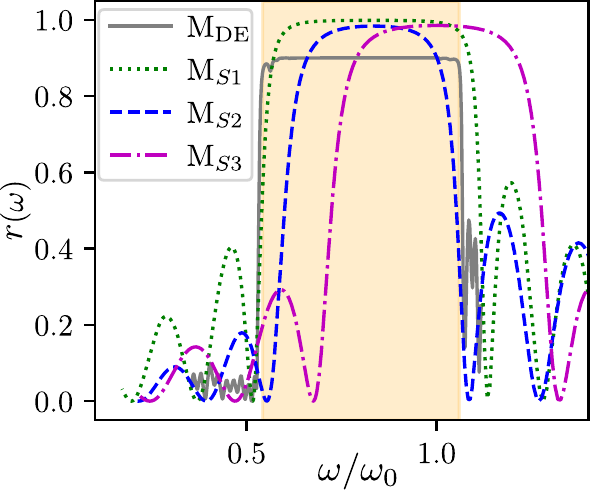}
    \caption{Calculated reflectivities for the structures. The laser frequency $\omega_0$ is the Nd:YAG frequency $\omega_0=2\pi\times $\SI{293.0}{THz}. The yellow-shaded region is the band of maximal reflectance of the $\mathrm{M}_{DE}$, that served as a guide for the optimization of the lightsail reflectors $\mathrm{M}_{S1}$ and $\mathrm{M}_{S2}$, whereas the reflector $\mathrm{M}_{S3}$ is designed to maximize the reflectance at $\omega=\omega_0$.}
    \label{fig:reflex}
\end{figure}

In order to achieve the required performances in the selected case, the design for $\mathrm{M}_{DE}$ is optimized using 3 stacks of silicon dioxide/titanium dioxide. Notice that the laser apertures are taken into account by weighting the overall reflectivity by a factor $b$, that is the ratio of the effective mirror surface to the total surface area at the array. This factor is conservatively set to $b=0.9$.
In the context of interstellar flight, the design of $\mathrm{M}_S$ is constrained by the requirement for a low surface mass density. Materials that were shown to be remarkably suitable for light sails
are titanium dioxide (TiO$_2$) \cite{santi2022multilayers} and silicon nitride (Si$_3$N$_4$) \cite{tung2022low} due to their relatively high index of refraction, high mid-infrared absorptivity, and thermal stability. 
In the present work, three examples of reflective multilayer sails are considered. They are named $\mathrm{M}_{S1}$, $\mathrm{M}_{S2}$, and $\mathrm{M}_{S3}$, consisting of Bragg reflectors. A multilayer reflector ($\mathrm{M}_{S3}$) is taken from a previous design of Tung and Davoyan
\cite{tung2022low}. It provides maximum reflectance at about \SI{1064}{nm}, therefore it is suitable for a DEP, but it turns out to be suboptimal for MDEP due to poor low-frequency performances. This design is evaluated along with two better performing structures, $\mathrm{M}_{S1}$ and $\mathrm{M}_{S2}$. $\mathrm{M}_{S2}$ is a multilayer that is based on $\mathrm{M}_{S3}$, with thicknesses optimized to achieve extended reflectance across the useful spectral range. $\mathrm{M}_{S1}$ is another optimized structure in which silicon nitride is substituted with titanium dioxide. The thickness of the reflectors is determined by optimizing it with IMD software \cite{windt1998imd} using a genetic algorithm.
The designs are represented in Fig.~\ref{fig:whereami}, and their corresponding reflectivities are reported in Fig.~\ref{fig:reflex}, with thicknesses reported in Table \ref{tab:designs} for the lightsail.
The $\mathrm{M}_{DE}$ mirror is a 30-layer stack made of SiO$_2$-TiO$_2$ is designed, in a configuration of 3 aperiodic stacks made of 10 layers, with thicknesses reported in the repository \cite{lorenzi2024}.
Surface densities are computed neglecting the presence of an additional structure on the back of the sail to facilitate the sail cooling, and density data are taken from Ref.~\cite{tpsx}.
\begin{table}
    \centering
    \begin{tabular}{l|c|l}
    \hline
    \hline
    \textbf{Parameter} & \textbf{Symbol} & \textbf{Value} \\
    \hline
    Lightsail mass & $m_S$ & \SI{10}{g}\\
    Average lightsail surface density & $\sigma$ & \SI{3e-3}{kg \per m^2} \\
    Sail diameter & $d_{S}$ & \SI{2.1}{m} \\
    Laser system power & $P_0$ & \SI{50}{GW} \\
    Laser system diameter & $d_\mathrm{DE}$ & \SI{3e4}{m} \\
    Laser wavelength & $\lambda_0$ & \SI{1064}{nm}\\
    Initial distance from laser system & $q_0$ & \SI{3.5e4}{km} \\
    Thrusting time & $t_F$ & \SI{500}{s}\\
    \hline
    \hline
    \end{tabular}
    \caption{Parameters utilized in the simulations.}
    \label{parameters}
\end{table}

In order to verity the actual performances of the designed cavities, the model has been applied to specific launch protocols, with parameters detailed in Table~\ref{parameters}. A circular sail of diameter $d_s$ is assumed, where the diameter is determined by the mass $\mathrm{M}_{S}$ and the uniform mass density $\sigma$ of the lightsail according to the relation $d_s = 2\sqrt{m_S/(\sigma\pi)}$. 
The average lightsail surface density used in simulation is \SI{3}{g/m^2}, which accounts for the additional weight of structures required for ligthsail stabilization \cite{tung2022low}. The material density data are taken from \cite{tpsx}. The performance of the reflective structures is evaluated computing $\Delta v$, representing the final velocity achieved after a constant thrust period, with the initial velocity set to zero. 
In the present case, the starting position $q_0$, as reported in Table~\ref{parameters}, is close to $\mathrm{M}_{DE}$, so there is a large number of reflections before the end of the thrusting period, as in the case reported in Fig.~\ref{main-point}b. 

\begin{figure}
    \centering
    \includegraphics[width=\columnwidth]{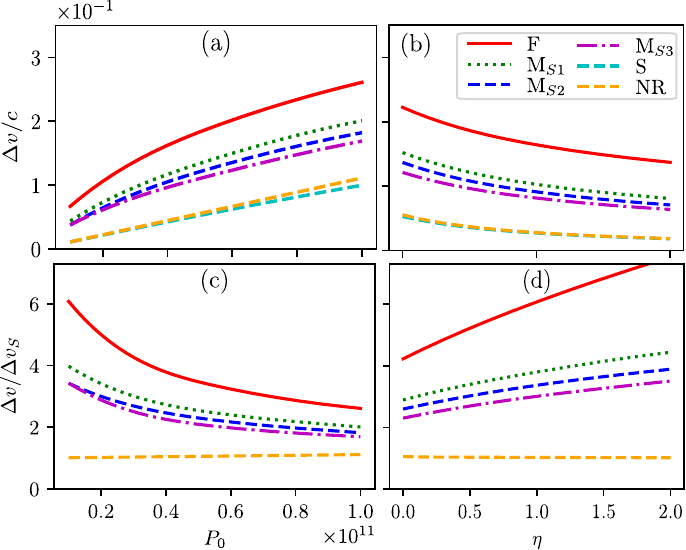}
    \caption{Final velocity $\Delta v$ resulting from the specified launch protocol as a function of the
    laser power $P_0$ (panel a and c), and the payload to sail mass ratio $\eta = m_p/m_{S}$ (panel b and d). 
    The value of the velocity normalized to the speed of light $\Delta v /c$ is shown in panels (a) and (b), while the ratio of the velocity with respect to the single-reflection result $\Delta v / \Delta v_S$ is shown in panels (c) and (d). 
    Panel (a) and (c) assumes a null ratio $\eta$, and panel (b) and (d) assumes a fixed power of \SI{50}{GW}.
    \Ra{The figures consider a span of parameters equal to one order of magnitude in $P_0$ and a factor of $2$ in $\eta$.}
    }
    \label{FoMo}
\end{figure}

In Fig.~\ref{FoMo} the final velocity is evaluated for the different multilayers, as a function of the payload-to-sail mass ratio $\eta = m_p/m_{S}$, and of the laser power $P_0$. The results when using a DEP and the nonrelativistic DEP model are also reported, according to the formulation of Tung and Davoyan~\cite{tung2022low}, which has been shown to predict slightly higher results in the relativistic regime. The effect of diffraction is also taken into account with the simplified model included in Eqs.~(\ref{eq:gamma-minestra}-\ref{eq:diffraction}). The simulation results are shown in Fig.~\ref{FoMo}.
Panels (a) and (b) show the final velocity $\Delta v$ as a function of the laser power $P_0$ and the mass ratio $\eta$. They show that the performance of the wideband structures is better than that of the structure $\mathrm{M}_{S3}$, optimized for the single reflection system. Panels (c) and (d) represent the ratio between the final velocity $\Delta v$ of the multiple reflection and nonrelativistic simulations with respect to the case of a single reflection $\Delta v_S$. Multiple reflections have a significant impact on the final velocity, especially for lowest values of the power.
\Ra{
Thermal stability is not directly addressed in this case, in order to focus on the relativistic dynamics predicted by the mechanical model, by considering a very lightweight lightsail. A simultaneous optimization of the thermal and mechanical parts needs to be developed in a more complete implementative study. 
All the simulations are performed assuming perfectly aligned mirrors and perfect fabrication.
However, the mirror roughness and imperfections are modeled by considering a variation of $5\%$ of the reflectivity of each mirror in the system. The resulting final velocity is changed by a maximum relative variation of less than $10\%$ in all the cases of all the structures in Table \ref{tab:designs}.
For what concerns the misalignment of the sail, other studies discuss techniques to stabilize the lightsail \cite{savu2022structural, Ilic2019, myilswamy2020photonic} and to control its orientational asset, paving the way for a structural control of the alignment. 
A deeper analysis will be needed in the implementation phase of the system.
}

\section{Summary}
The photon recycling system proposed represents a compelling strategy for increasing the acceleration in reflective lightsails, and it is particularly suited to achieving relativistic speeds required in interstellar exploration projects. A theoretical model that takes into account the delay due to the finite propagation velocity of light and the relativistic Doppler frequency shift has been developed to improve the past estimates on the power delivered with such a system. More in general, the proposed theory solves the problem of the electromagnetic field in an optical resonator with a movable mirror in relativistic regime. 
A remarkable outcome of the application of this model is that the requirements on the idealized reflectors in the cavity are independent on launch parameters, so that the same system can be re-used for different launch scenarios. The mirrors must efficiently reflect frequencies above a set threshold, which contribute significatively to the propulsion, while filtering the far-shifted waves that fall inside the band of thermal emission of the lightsail. By imposing such requirements on the spectral response of the mirrors, some designs have been proposed and their performance compared by simulating the thrusting with realistic lightsail parameters. Results show that some optimized designs are more effective than others and quantify the enhanced performance of the system compared to a single-reflection laser propulsion system.

\section*{Acknowledgements}
The authors thank A. J. Corso and A. J. Higgins for useful scientific discussion.

\section*{Appendices}
\subsection{Relativistic Newton equation}
The Newton equation is obtained in the relativistic case starting from the definition of relativistic momentum $\mathbf{p}=\gamma m \mathbf{v}$, we can write
\begin{equation}\label{eq:relentum}
    \dv{\mathbf{p}}{t} = \dv{\gamma}{t} m \mathbf{v} + \gamma m \dv{\mathbf{v}}{t},
\end{equation}
by defining the longitudinal and transverse acceleration, with respect to the vector $\mathbf{v}$ 
\begin{equation}
    \dv{\mathbf{v}}{t} = \mathbf{a}_L +\mathbf{a}_T \,,
\end{equation}
it is simple to show that the first term in Eq.~(\ref{eq:relentum}) is only longitudinal 
\begin{equation}
    \dv{\gamma}{t} m \mathbf{v} = \frac{v^2}{c^2} \gamma^3 m \mathbf{a}_L \,,
\end{equation}
and the second term is composed of longitudinal and transverse parts
\begin{equation}
    \gamma m \dv{\mathbf{v}}{t}=\gamma m (\mathbf{a}_L+\mathbf{a}_T) \,.
\end{equation}
Summing the contributions, we can separate the Newton equation with respect to transverse and longitudinal forces, obtaining
\begin{equation}
    \dv{\mathbf{p}}{t}=\gamma^3m \mathbf{a}_L+\gamma m \mathbf{a}_T \,.
\end{equation}
In the absence of transverse force, like in the case of radiation pressure from a longitudinal beam Eq.~(\ref{eq:newton}) is retrieved.

\Ra{

\subsection{Derivation of the relativistic Doppler coefficient}
The derivation of the relativistic Doppler effect is formally achieved by considering the Lorentz transformation of the plane wave $4$-vector, that we may represent by $K^\mu = (\omega/c, \mathbf{k})$, taking $\mathbf{k}=k\mathbf{u}_x$ aligned with the $x$ axis. Considering a Lorentz boost in the $x$ (longitudinal) direction, the transformation gives $\tilde{\omega}=\gamma(\omega-vk)$ and $\tilde{k}=\gamma(k-v \ \omega/c^2)$. By using the dispersion relation of the free plane wave, $k=\omega/c$, we get 
\begin{equation}
\frac{\tilde{\omega}}{\omega}=\gamma(1-\beta) = \sqrt{\frac{1-\beta}{1+\beta}} \,.
\end{equation}
In a process where the wave is reflected back along the same direction, the transformation applies twice. This can be seen by considering the reflected wave as a new wave, undergoing the same Doppler shift. Therefore the coefficient $D$ we use in Eq.~(\ref{coffi}) is the square of the coefficient of the above equation.
We use $D$ in most of the study. However, in Eq.~(\ref{welativistico}), that is a relation where it is necessary to consider the frequency in the lightsail reference frame, the value $\sqrt{D}$ is used instead.

An interesting property of the relativistic Doppler effect in the longitudinal direction is that it can be derived simply by correcting the classical Doppler effect.
The classical Doppler formula for light would read $\omega' = \omega/(1+\beta)$, where the prime indicates the change of reference system in the classical limit. The transformation of the time coordinate is $\tilde{t}=\gamma t$, and by correcting the classical value as $\tilde{\omega}=\omega'\gamma^{-1}$, we get the exact value of the relativistic Doppler frequency.
}

\bibliography{prapplied}

\end{document}

%% file: commands.tex
\usepackage{siunitx}
\usepackage{amsmath,amssymb,mathrsfs}
\usepackage{psfrag}
\usepackage{graphicx}
\usepackage{graphics}
\usepackage{epsfig}
\usepackage{bm}
\usepackage{color}
\usepackage{verbatim,color}
\usepackage{physics}
\usepackage[colorlinks=true,linkcolor=black,citecolor=black,urlcolor=black]{hyperref}

\usepackage[acronym]{glossaries}
\newacronym{dr}{DR}{dimensional regularization}
\newacronym{ms}{MS}{minimal subtraction}
\newacronym{eft}{EFT}{effective field theory}
\newacronym{mqst}{MQST}{macroscopic quantum self-trapping }
\newacronym{lhy}{LHY}{Lee-Huang-Yang}
\newcommand{\beq}{\begin{equation}}
\newcommand{\eeq}{\end{equation}}
\newcommand{\beqa}{\begin{eqnarray}}
\newcommand{\eeqa}{\end{eqnarray}}
\newcommand{\ba}{\begin{aligned}[b]}
\newcommand{\ea}{\end{aligned}}

\definecolor{darkgreen}{rgb}{0.0, 0.5, 0.0}
\definecolor{shadow}{rgb}{0.6, 0.6, 0.6}
\definecolor{darkgreen}{rgb}{0.0, 0.5, 0.0}
\definecolor{c1}{rgb}{0.9, 0.0, 0.0}
\definecolor{c2}{rgb}{0.4, 0.0, 0.7}
\definecolor{c3}{rgb}{0.9, 0.0, 0.0}
\newcommand{\Ra}[1]{\color{black}#1\color{black}}
\newcommand{\Rb}[1]{\color{black}#1\color{black}}

\usepackage{ulem}